\documentclass[12pt]{article}
  \usepackage{amsfonts}
  \usepackage{amsmath}
\usepackage{amssymb}
\usepackage{amscd}
\usepackage[dvips]{graphicx}
\begin{document}
~~
\bigskip
\bigskip
\begin{center}
%\section*
{\Large {\bf{{{Canonical and Lie-algebraic twist deformations of
Galilei algebra}}}}}
\end{center}
\bigskip
\bigskip
\bigskip
\begin{center}
{{\large ${\rm {Marcin\;Daszkiewicz}}$ }}
\end{center}
\bigskip
\begin{center}
{ {{{Institute of Theoretical Physics\\ University of Wroc{\l}aw pl.
Maxa Borna 9, 50-206 Wroc{\l}aw, Poland\\ e-mail:
marcin@ift.uni.wroc.pl}}}}
\end{center}
\bigskip
\bigskip
\bigskip
\bigskip
\bigskip
\bigskip
\bigskip
\bigskip
\begin{abstract}
We describe various nonrelativistic contractions of two classes of
twisted Poincare algebra: canonical one
($\theta_{\mu\nu}$-deformation) and the one leading to Lie-algebraic
models of noncommutative space-times. The cases of
contraction-independent and contraction-dependent twist parameters
are considered. We obtain five models of noncommutative
nonrelativistic space-times, in particular, two new Lie-algebraic
nonrelativistic deformations of space-time, respectively, with
quantum time/classical space and with quantum space/classical time.
\end{abstract}
\bigskip
\bigskip
\bigskip
\bigskip
\eject
\section{{{Introduction}}}

In the last decade the interest in a class of theories describing
kinematics  different from  Special Relativity is growing rapidly.
The main reason for such type of considerations follows from many
phenomenological suggestions, which state that  relativistic
space-time symmetries should be modified (deformed) at Planck scale,
while  the classical Poincare invariance still remains  valid at
larger distances \cite{1a}-\cite{1d}. There are also several formal
arguments based mainly on quantum gravity \cite{2}, \cite{2a} and
string theory \cite{1s}, \cite{2s} indicating that space-time at
Planck-length should be noncommutative, i.e. it should have a
quantum nature.

An important candidate for a modification of relativistic symmetry
in ultra-high energy regime, so-called $\kappa$-deformed
Poincar\'{e} algebra, has been proposed in the Hopf algebraic
framework of quantum groups in  \cite{4a}, \cite{4b}. As a result of
 contraction  of q-deformed anti-De-Sitter algebra one
obtains $\kappa$-Poincar\'{e} group which leads to   Lie-algebraic
space-time noncommutativity, the $\kappa$-Minkowski space
\cite{min}, \cite{min1}, with the mass-like deformation parameter
$\kappa$.
%usually linked with  the Planck mass.
Besides, it also gives   a formal framework for such theoretical
constructions as Double Special Relativity (see e.g.
\cite{dsr1a}-\cite{dsr1d}), which postulates two
observer-independent scales, of velocity, describing the speed of
light, and of mass, which can be identify with $\kappa$-parameter
and is expected to be of the order of fundamental Planck mass.

In accordance with the general classification of all possible
deformations of the Poincar\'{e} group \cite{zakrzewski},  there
exist another two interesting modifications of relativistic
symmetries. First of them corresponds to the so-called soft twisted
Poincar\'{e} Hopf algebra, and leads to very popular at present,
canonical commutation relations for quantum Minkowski space
\cite{3e}-\cite{3d}
\begin{equation}
[\;{\hat x}_{\mu},{\hat x}_{\nu}\;] = i\theta_{\mu\nu}\;\;\;;\;\;\;
\theta_{\mu\nu} = {\rm const}\;. \label{noncomm}
\end{equation}
The second deformation, associated with  deformation as well
generated by twist, is closer to the $\kappa$-deformed case as it
introduces the Lie-algebraic type of space-time noncommutativity
(\cite{lie2}; see also \cite{lie1})
\begin{equation}
[\;{\hat x}_{\mu},{\hat x}_{\nu}\;] =
i\theta_{\mu\nu}^{\rho}x_{\rho}\;, \label{noncomm1}
\end{equation}
with particularly chosen coefficients $\theta_{\mu\nu}^{\rho}$ being
constants. Nevertheless, it should be noted that such a
Lie-algebraic  modification of relativistic symmetry looks simpler
than the  $\kappa$-deformed one.
% from formal point of view.
Its classical r-matrix satisfies the classical Yang-Baxter equation,
and  it can be generated  by a twist transformation of the
undeformed Poincar\'{e} Hopf structure \cite{twist}. Besides, in
contrast to the $\kappa$-deformed case, the explicit form of the
corresponding quantum R-matrix is  known.

The   deformations (\ref{noncomm}) and (\ref{noncomm1}) are supposed
to describe the kinematics of a very fast (relativistic) object in
 ultra-high energy regime. However, one can
ask about their nonrelativistic limit, i.e. about their deformed
Galilei counterparts describing much  slower objects in the
transplanckian region. In the case of $\kappa$-Poincar\'{e} algebra
there exist two   nonrelativistic contractions to Galilei group
depending on the way in which we embed parameter $c$ in $\kappa$
\cite{kappaga}-\cite{azca2}. First contraction uses substitution
$\kappa = {\hat{\kappa}}/c$ and in the contraction limit $c\to
\infty$ follows  the $\kappa$-deformed Galilei algebra firstly
studied in the article \cite{kappaga}\footnote{This contraction is
rather of mathematical nature. If we relate e.g. $\kappa$ with
Planck mass, this contraction corresponds to vanishing Planck
mass.}. The second one ($\kappa$ is replaced by ${\hat{\kappa}}c$)
has been investigated in the context of so-called extended deformed
Galilei Hopf structure (\cite{azca1}, \cite{azca2}, see also
\cite{azca3}-\cite{ball}), and in the presence of additional central
generator leads to the extended $\kappa$-deformed Galilei group
\cite{azca1}. The main consequences of these contracted algebras,
for example the type of corresponding noncommutative space-times or
differential calculi, have been studied in  \cite{azca1},
\cite{azca2}. Besides, it was also shown (see e.g. \cite{gg}) that
the above quantum groups, similarly as their relativistic
counterparts, exhibit the bicross-product structure \cite{majid}.

The main aim of  present paper is to consider nonrelativistic
quantum contractions with $c$-dependent
 twist parameters in context of  canonical and
Lie-algebraic twisted Poincar\'{e} group \cite{3a}, \cite{lie2}. In
such a way we find three new, deformed Galilei Hopf algebras, which
can be also obtained by twisting of the classical Galilei Hopf
algebras. Further, we also investigate the simplest type of
relativistic contraction with $c$-independent deformation parameter
\cite{inonu}
 (see also \cite{barut}). In such a way we recover two
quantum groups corresponding to the soft- and Lie-twisted Poincare
algebras - the soft- and Lie-twisted Galilei algebras. Finally, we
derive the corresponding noncommutative "space-times" describing the
representation spaces (Hopf modules) of our Galilean twisted quantum
groups \cite{bloch}, \cite{3a}, \cite{3b}.

The paper is organized as follows. In second Section we recall basic
facts concerning the soft and Lie-algebraic twisted relativistic
symmetries. Sections three and four are devoted to their
nonrelativistic contractions to Galilei quantum groups with
contraction-independent and contraction-dependent twist parameter,
respectively. In Section five we derive the corresponding
nonrelativistic deformed space-times. The results are discussed and
summarized in the last Section.

\section{Canonical and Lie-algebraic twisted Poincar\'{e} algebra}

In this section we review some known facts concerning the soft and
Lie-algebraic twisted classical Poincar\'{e} algebras (see
\cite{3a}, \cite{3b},  \cite{lie2}), which we denote by $\,{\mathcal
U}_{\theta}({\mathcal P})$ and $\,{\mathcal U}_{\zeta}({\mathcal
P})$, respectively.

First of all, let us start with canonical twisted modification of
relativistic symmetry with carrier algebra described by commuting
fourmomentum generators $P_\mu$. According to \cite{zakrzewski} (see
also \cite{3e}) its classical $r$-matrix looks as follows
\begin{equation}
r_{\theta}=\frac{1}{2}\theta ^{\mu \nu }\,P_{\mu }\wedge P_{\nu }\;,
\label{1}
\end{equation}
with $a\wedge b=a\otimes b-b\otimes a$ and $\theta^{\mu\nu} =
-\theta^{\nu\mu}$. Such Abelian  $r$-operator satisfies trivially
the classical Yang-Baxter equation (CYBE)
\begin{equation}
[[\;r_{\theta},r_{\theta}\;]] = [\;r_{\theta12},r_{\theta13} +
r_{\theta23}\;] + [\;r_{\theta13}, r_{\theta23}\;] = 0\;,
\label{cybe}
\end{equation}
where  the  symbol $[[\;\cdot,\cdot\;]]$ denotes the Schouten
bracket while $r_{\theta 12} = \frac{1}{2}\theta ^{\mu \nu }P_{\mu
}\wedge P_\nu\wedge 1$, $r_{\theta 13} = \frac{1}{2}\theta ^{\mu \nu
}P_{\mu }\wedge 1 \wedge P_\nu$ and $r_{\theta 23} =
\frac{1}{2}\theta ^{\mu \nu }1\wedge P_\mu\wedge P_\nu$. We get the
$\theta^{\mu\nu}$-deformed Poincar\'{e} algebra  by twisting of the
classical coproducts with use of the  twist factor
\begin{equation}
\mathcal{F}_{\theta }=\exp \frac{i}{2}(\theta ^{\mu \nu }P_{\mu
}\wedge P_{\nu }\,)\;,  \label{factor}
\end{equation}
satisfying the classical cocycle condition \cite{drin}
\begin{equation}
{\mathcal F}_{{\theta }12} \cdot(\Delta_{0} \otimes 1) ~{\cal
F}_{\theta } = {\mathcal F}_{{\theta }23} \cdot(1\otimes \Delta_{0})
~{\mathcal F}_{{\theta }}\;, \label{cocyclef}
\end{equation}
and the normalization one
\begin{equation}
(\epsilon \otimes 1)~{\cal F}_{{\theta }} = (1 \otimes
\epsilon)~{\cal F}_{{\theta }} = 1\;, \label{normalizationhh}
\end{equation}
with ${\cal F}_{{\theta }12} = {\cal F}_{{\theta }}\otimes 1$ and
${\cal F}_{{\theta }23} = 1 \otimes {\cal F}_{{\theta }}$. The
algebraic sector of $\,{\mathcal U}_{\theta}({\mathcal P})$ is not
changed, while  coproducts and antipodes transform according to
\begin{equation}
\Delta _{0}(a) \to \Delta _{\theta }(a) = \mathcal{F}_{\theta }\circ
\,\Delta _{0}(a)\,\circ \mathcal{F}_{\theta }^{-1}\;,\label{fc}
\end{equation}
\begin{equation}
S_{\theta}(a) =u(\theta)\,S_{0}(a)\,u^{-1}(\theta)\;,\label{fs}
\end{equation}
where $\Delta _{0}(a) = a\otimes 1+ 1\otimes a$, $S_0(a) = -a$ and
$u(\theta)=\sum f_{(1)}S_0(f_{(2)})$ (we use Sweedler's notation
$\mathcal{F}=\sum f_{(1)}\otimes f_{(2)}$). The twisted Hopf algebra
$\,{\mathcal U}_{\theta}({\mathcal P})$ has the form ($\eta_{\mu\nu}
= (-,+,+,+)$)
\begin{eqnarray}
&&\left[ \,M_{\mu \nu },M_{\rho \sigma }\,\right] =i\left( \eta
_{\mu \sigma }\,M_{\nu \rho }-\eta _{\nu \sigma }\,M_{\mu \rho
}+\eta _{\nu \rho }M_{\mu
\sigma }-\eta _{\mu \rho }M_{\nu \sigma }\right) \;,  \label{dlww2.1x} \\
&~~&  \cr &&\left[ \,M_{\mu \nu },P_{\rho }\,\right] =i\left( \eta
_{\nu \rho }\,P_{\mu }-\eta _{\mu \rho }\,P_{\nu }\right)
\;\;\;,\;\;\;\left[ \, P_{\mu },P_{\nu }\,\right] =0\;, \notag
\end{eqnarray}
with coalgebra
\begin{eqnarray}
\Delta_\theta(P_\mu)&=&\Delta_0(P_\mu)\;, \nonumber\\
&~~&  \cr \Delta _{\theta }(M_{\mu \nu })&=&\mathcal{F}_{\theta
}\circ \,\Delta _{0}(M_{\mu \nu
})\circ \mathcal{F}_{\theta }^{-1}\nonumber\\
&=&\Delta _{0}(M_{\mu \nu })-%
\theta ^{\rho \sigma }[(\eta _{\rho \mu }P_{\nu }-\eta _{\rho \nu
}\,P_{\mu })\otimes P_{\sigma }\label{dlww33}\\
&&\qquad\qquad\qquad\qquad\qquad+P_{\rho}\otimes (\eta_{\sigma
\mu}P_{\nu}-\eta_{\sigma \nu}P_{\mu})]\;,\nonumber
\end{eqnarray}
and the classical antipodes and counits
\begin{equation}
\qquad S_{\theta}(P_{\mu })=-P_{\mu }\;\;\;,\;\;\;S_{\theta}(M_{\mu
\nu })=-M_{\mu \nu }\;\;\;,\;\;\;  %\label{dlww2.23}
\epsilon(M_{\mu \nu })=\epsilon(P_\mu)=0\;. \label{counit}
\end{equation}

Let us now turn to the Lie-algebraic twist of Poincare algebra
$\,{\mathcal U}_{\zeta}({\mathcal P})$ \cite{lie2} (see also
\cite{lie1}) with carrier algebra described by three commuting
generators $M_{\alpha\beta}\,$, $P_{\epsilon }$ where $\epsilon \neq
\alpha,\,\beta$ and where the indices $\alpha,\,\beta$ are fixed.
The corresponding classical $r$-matrix has the form
\cite{zakrzewski}
\begin{equation}
r_\zeta= \frac{1}{2}\zeta^\lambda\,P_{\lambda }\wedge M_{\alpha
\beta } \;, \label{dlww11}
\end{equation}
with the vector $\zeta^\lambda$ having vanishing components
$\zeta^\alpha,\ \zeta^\beta$.
In particular, the $r$-matrix (\ref{dlww11}) for the special choices $%
\{M_{\alpha\beta}= M_{03},\epsilon=1,2\}$ and $%
\{M_{\alpha\beta}=M_{12},\epsilon=0,3\}$ has been discussed some
time ago  in \cite{lie1} and also considered very recently in
\cite{lala}. Obviously, the operator (\ref{dlww11}) satisfies the
classical Yang-Baxter equation (\ref{cybe}), and twist factor
corresponding to (\ref{dlww11}) looks as follows
\begin{equation}
\mathcal{F}_{\zeta}=  \exp \,\frac{i}{2}( \zeta^\lambda\,P_{\lambda
}\wedge M_{\alpha \beta }) \;. \label{swarog}
\end{equation}
Using the twist procedure one gets (besides the undeformed algebraic
commutation relations (\ref{dlww2.1x})) the following coproducts
\cite{lie2}
\begin{eqnarray}
 \Delta_\zeta(P_\mu)&=&\Delta
_0(P_\mu)+(-i)^\gamma\sinh(i^{\gamma} \xi^\lambda  P_\lambda )\wedge
\left(\eta_{\alpha \mu}P_\beta -\eta_{\beta \mu}P_\alpha \right)\label{dlww2.2}\\
&+&(\cosh(i^{\gamma}\xi^\lambda  P_\lambda )-1)\perp
\left(\eta_{\alpha \alpha }\eta_{\alpha \mu}P_\alpha +\eta_{\beta
\beta }\eta_{\beta \mu}P_\beta \right)\;,
\notag  \\
&~~&  \cr \Delta_\zeta(M_{\mu\nu})&=&\Delta_0(M_{\mu\nu})+M_{\alpha
\beta }\wedge \xi^\lambda \left(\eta_{\mu
\lambda }P_\nu-\eta_{\nu \lambda }P_\mu\right)\label{dlww2.22}\\
&+&i\left[M_{\mu\nu},M_{\alpha \beta }\right]\wedge
(-i)^{\gamma}\sinh(i^{\gamma}\xi^\lambda P_\lambda ) \notag \\
&+&\left[\left[%
M_{\mu\nu},M_{\alpha \beta }\right],M_{\alpha \beta
}\right]\perp(-1)^{1+\gamma}
(\cosh(i^{\gamma}\xi^\lambda  P_\lambda  )-1)  \nonumber \\
&+&M_{\alpha \beta }(-i)^\gamma\sinh(i^{\gamma}\xi^\lambda P_\lambda
)\perp
\xi^\lambda \left(\psi_\lambda P_\alpha -\chi_\lambda P_\beta \right) \notag \\
&+&\xi^\lambda \left(\psi_\lambda \eta_{\alpha \alpha }P_\beta
+\chi_\lambda \eta_{\beta \beta }P_\alpha \right)\wedge M_{\alpha
\beta }(-1)^{1+\gamma}(\cosh(i^{\gamma}\xi^\lambda P_\lambda )-1)\;,
  \notag
\end{eqnarray}
where, in the above formulas $a\perp b=a\otimes b+b\otimes a$,
\begin{equation*}
\xi^\lambda=\frac{\zeta^\lambda}{2}\,,\qquad \psi_\lambda =\eta_{\nu
\lambda }\eta_{\beta \mu}-\eta_{\mu \lambda }\eta_{\beta
\nu}\,,\qquad \chi_\lambda =\eta_{\nu \lambda }\eta_{\alpha
\mu}-\eta_{\mu \lambda }\eta_{\alpha \nu}\,,
\end{equation*}
and $\gamma=0$ when $M_{\alpha\beta}$ is a boost or $\gamma=1$ for a
space rotation. Antipodes and counits remain classical.
%\begin{equation}
%\qquad S_{\zeta}(P_{\mu })=-P_{\mu }\;\;\;,\;\;\;S_{\zeta}(M_{\mu
%\nu })=-M_{\mu \nu }\,,  \label{dlww2.23gg}
%\end{equation}
%\begin{equation}
%\epsilon(M_{\mu \nu })=\epsilon(P_\mu)=0\,. \label{counitss}
%\end{equation}
The relations (\ref{dlww2.1x}) and (\ref{dlww2.2}), (\ref{dlww2.22})
define the Lie-algebraic twist deformation $\,{\mathcal
U}_{\zeta}({\mathcal P})$ of  Poincar\'{e} algebra. Of course, we
get the undeformed classical Poincar\'{e} Hopf structure
$\,{\mathcal U}_{0}({\mathcal P})$ in the $\zeta \to 0$ limit.

\section{Nonrelativistic contractions of twisted Poincar\'{e} algebras:
contraction-independent twist parameters}

In this section we present the  nonrelativistic contractions of Hopf
structures described in previous section, i.e. we find their
nonrelativistic counterparts - the $\theta^{\mu\nu}$-Galilei algebra
$\,{\mathcal U}_{\theta}({\mathcal G})$ and Lie-algebraic one
$\,{\mathcal U}_{\zeta}({\mathcal G})$. For this purpose let us
introduce the following standard redefinition of Poincar\'{e}
generators \cite{inonu}
\begin{equation}
P_{0 } = \frac{{\Pi_{0 }}}{c}\;\;\;,\;\;\;P_{i } =
\Pi_{i}\;\;\;,\;\;\;M_{ij}= K_{ij}\;\;\;,\;\;\;M_{i0}= cV_i\;,
\label{contr2}
\end{equation}
where parameter $c$ describes the light velocity. We begin with
canonical twisted algebra $\,{\mathcal U}_{\theta}({\mathcal P})$.
In a first step of contraction procedure one can rewrite the
algebraic relations (\ref{dlww2.1x}), coproducts (\ref{dlww33}),
antipodes %(\ref{dlww2.23})
and counits (\ref{counit}) in terms of new generators $\Pi_{0 }$,
$\Pi_{i}$, $K_{ij}$, $V_i$. Next, one should take a proper
nonrelativistic limit $(c\to \infty)$ of rewritten formulas. In such
a way we obtain the classical $D=4$ Galilean algebra
\begin{eqnarray}
&&\left[\, K_{ij},K_{kl}\,\right] =i\left( \delta
_{il}\,K_{jk}-\delta
_{jl}\,K_{ik}+\delta _{jk}K_{il}-\delta _{ik}K_{jl}\right) \;,  \notag \\
&~~&  \cr &&\left[\, K_{ij},V_{k}\,\right] =i\left( \delta
_{jk}\,V_i-\delta _{ik}\,V_j\right)\;\; \;, \;\;\;\left[
\,K_{ij},\Pi_{\rho }\right] =i\left( \eta _{j \rho }\,\Pi_{i }-\eta
_{i\rho }\,\Pi_{j }\right) \;, \label{nnn}
\\
&~~&  \cr &&\left[ \,V_i,V_j\,\right] = \left[ \,V_i,\Pi_{j
}\,\right] =0\;\;\;,\;\;\;\left[ \,V_i,\Pi_{0 }\,\right]
=-i\Pi_i\;\;\;,\;\;\;\left[ \,\Pi_{\rho },\Pi_{\sigma }\,\right] =
0\;,\nonumber
\end{eqnarray}
and the following deformation of coalgebraic sector
\begin{eqnarray}
&&\Delta_\theta(\Pi_\rho)=\Delta_0(\Pi_\rho)\;\;\;,\;\;\;
\Delta _{\theta }(V_i) =\Delta _{0}(V_i)\;, \label{dlww3v}\\
&~~~&  \cr \Delta _{\theta }(K_{ij})
&=&\Delta _{0}(K_{ij})-%
\theta ^{k l }[(\delta_{k i}\Pi_{j }-\delta_{k j
}\,\Pi_{i})\otimes \Pi_{l }\\
&&\qquad\qquad\qquad\qquad\qquad+\Pi_{k}\otimes (\delta_{l
i}\Pi_{j}-\delta_{l j}\Pi_{i})]\;.\label{zadruzny}
\end{eqnarray}
The antipodes and counits remain undeformed
\begin{equation}
 S(\Pi_{\rho})=-\Pi_{\rho}\;\;\;,\;\;\;S(K_{ij})=-K_{ij}\;\;\;,\;\;\;S(V_i)=-V_i\;,  \label{zadruga}
\end{equation}
\begin{equation}
 \epsilon(\Pi_{\rho}) = \epsilon(K_{ij}) = \epsilon(V_i) = 0\;,  \label{zadruga1}
 \vspace{0.3cm}
\end{equation}
where $k,l = 1,2,3$. The relations (\ref{nnn})-(\ref{zadruga1})
constitute the Hopf algebra structure which we shall call the
canonically  deformed Galilei Hopf algebra $\,{\mathcal
U}_{\theta}({\mathcal G})$.  We note that one can also perform the
contraction of  classical $r$-matrix (\ref{1}) and of the twist
factor (\ref{factor}). We get (\cite{kowclas}, see also
\cite{lukworproc})
\begin{equation}
\texttt{r}_{\theta}=\frac{1}{2}\theta ^{kl}\,\Pi_{k}\wedge
\Pi_{l}\;, \label{grmatix}
\end{equation}
and
\begin{equation}
{\mathcal{K}}_{\theta }=\exp \frac{i}{2}(\theta ^{kl}\Pi_{k }\wedge
\Pi_{l})\;, \label{gfactor}
\end{equation}
respectively. We see that the $\texttt{r}$-operator (\ref{grmatix})
satisfies the classical Yang-Baxter equation (\ref{cybe}) while the
factor (\ref{factor}) defines  twist element, which satisfies the
classical cocycle and normalization conditions (\ref{cocyclef}),
(\ref{normalizationhh}), and provides the formulas
(\ref{dlww3v})-(\ref{zadruzny})
%Hence, we may get  $\,{\mathcal U}_{\theta}({\mathcal G})$ by
%twisting the classical (undeformed) Galilei algebra ($\,{\mathcal
%U}_{\theta = 0}({\mathcal G})$)  with help of the factor
%(\ref{gfactor}).

In the case of  Lie-type twisted Poincar\'{e}  algebra $\,{\mathcal
U}_{\zeta}({\mathcal P})$ the contraction is more complicated. For
carrier algebra $\{\,M_{kl}, P_\gamma\;;\; \gamma
\ne\,k,\,l,\,0\,\}$  one can check that after contracting the
formulas (\ref{dlww2.2}) and (\ref{dlww2.22}) we get the classical
Galilei  algebraic sector (\ref{nnn}) and the following deformed
coproducts
\begin{eqnarray}
 \Delta_\zeta(\Pi_0)&=&\Delta _0(\Pi_0)\;,\label{ffdlww2.2}\\
&~~&  \cr \Delta_\zeta(\Pi_i)&=&\Delta _0(\Pi_i)+\sin( \xi
\Pi_\gamma )\wedge
\left(\delta_{k i}\Pi_l -\delta_{l i}\Pi_k \right)\\
&+&(\cos(\xi  \Pi_\gamma )-1)\perp \left(\delta_{k i}\Pi_k
+\delta_{l i}\Pi_l \right)\;, \nonumber
\end{eqnarray}
\begin{eqnarray}
~~ \Delta_\zeta(K_{ij})&=&\Delta_0(K_{ij})+K_{k l }\wedge \xi
\left(\delta_{i
\gamma }\Pi_j-\delta_{j \gamma }\Pi_i\right)\nonumber\\
&+&i\left[K_{ij},K_{k l }\right]\wedge
\sin(\xi \Pi_\gamma ) \notag \\
&+&\left[\left[%
K_{ij},K_{k l }\right],K_{k l }\right]\perp
(\cos(\xi  \Pi_\gamma  )-1)  \label{ffdlww2.22} \\
&+&K_{k l }\sin(\xi \Pi_\gamma )\perp
\xi \left(\psi_\gamma \Pi_k -\chi_\gamma \Pi_l \right) \notag \\
&+&\xi \left(\psi_\gamma \Pi_l +\chi_\gamma \Pi_k \right)\wedge K_{k
l }(\cos(\xi \Pi_\gamma )-1)\;,
  \notag
\end{eqnarray}
\begin{eqnarray}
\Delta_\zeta(V_i)&=&\Delta_0(V_i) +i\left[V_i,K_{k l }\right]\wedge
\sin(\xi \Pi_\gamma )
\label{fifdlww2.22}\\
&+&\left[\left[%
V_i,K_{k l }\right],K_{k l }\right]\perp (\cos(\xi \Pi_\gamma
)-1)\;,  \nonumber
\end{eqnarray}
where
\begin{equation*}
\xi=\frac{\zeta}{2}\,,\qquad \psi_\gamma =\delta_{j \gamma
}\delta_{l i}-\delta_{i \gamma }\delta_{l j}\,,\qquad \chi_\gamma
=\delta_{j \gamma }\delta_{k i}-\delta_{i \gamma }\delta_{k j}\;.
\end{equation*}
The antipodes and counits remain classical (see (\ref{zadruga}),
(\ref{zadruga1})).  One can also derive the contraction limit for
the $\zeta$-deformed classical $\texttt{r}$-matrix
\begin{equation}
\texttt{r}_\zeta= \frac{1}{2}\zeta\,\Pi_{\gamma}\wedge K_{kl} \;,
\label{dlww11new}
\end{equation}
which satisfies  the classical Yang-Baxter equation
\begin{equation}
[[\;\texttt{r}_\zeta,\texttt{r}_\zeta\;]]  = 0\;. \label{cybenew}
\end{equation}
Similarly, by contracting (\ref{dlww11})  we get the following twist
factor
\begin{equation}
\mathcal{K}_{\zeta}=  {\rm \exp}
\,\frac{i}{2}(\zeta\,\Pi_{\gamma}\wedge K_{kl}) \;, \label{swarozyc}
\end{equation}
which can be used to get $\,{\mathcal U}_{\zeta}({\mathcal G})$ by
twisting the classical Galilei Hopf algebra $\,{\mathcal
U}_{0}({\mathcal G})$.

In the case $\{\,M_{kl}, P_0\;(\gamma = 0) \, \}$ the contracted
algebra is trivial, i.e. its coproducts are primitive while its
classical $\texttt{r}$-matrix disappears
\begin{equation}
\texttt{r}_\zeta= 0 \;\;\;,\;\;\;\Delta_{\zeta}(a) = \Delta_{0}
(a)\;.
\end{equation}
On the other side, for the carrier algebra containing boosts
$\{\,M_{k0}, P_l\;;\;k\ne l\,\}$ the contraction limit
(\ref{contr2}) does not exist. In fact, one can check that  the
coproducts for rotations and corresponding classical
$\texttt{r}$-matrix are divergent in the contraction limit $c \to
\infty$.

\section{Nonrelativistic contractions of twisted Poincar\'{e} algebras:
contraction-dependent twist parameters}

Let us now look at the contraction with light velocity $c$ hidden in
the deformation parameter of quantum group. As already mentioned in
Introduction such a type of contraction has been investigated in the
case of $\kappa$-Poincar\'{e} Hopf algebra  \cite{azca1},
\cite{azca2}.

We begin with the canonical deformation $\,{\mathcal
U}_{\theta}({\mathcal P})$ with $\theta^{\mu\nu} =
{\hat{\theta}^{\mu\nu}}/{\kappa}$
$(\left[\,{\kappa}\,\right]=L^{-1})$. Let us introduce the parameter
$\hat{\kappa}$ such that ${\kappa}=\hat{\kappa}/c$
$(\left[\,\hat{\kappa}\,\right]=T^{-1})$. We perform the contraction
limit of co-sector (\ref{dlww33}) in two steps. Firstly, we rewrite
the formula (\ref{dlww33})  in term of the operators (\ref{contr2})
and parameter $\hat{\kappa}$.  Next, we take the $c\to \infty$ limit
and in such a way, in the case $\theta^{ij} =0$ and $\theta^{0i} =
\frac{{\hat \theta}^{0i}c}{{\hat \kappa}}$, one gets
\begin{eqnarray}
 \lim_{c\to \infty}\;\Delta_{\theta^{0i}}(\Pi_\mu)&=&\Delta _0(\Pi_\mu)\;,\label{zcoppy1}\\
 &~~&  \cr
\lim_{c\to \infty}\;\Delta_{\theta^{0i}}(K_{ij})&=&\Delta_0(K_{ij})
- \frac{\hat{\theta}^{0k}}{\hat{\kappa}}\Pi_0
\wedge\left(\delta_{ki}\Pi_j - \delta_{kj}\Pi_i\right)
%-\frac{\hat{\theta}^{0k}}{\hat{\kappa}} \left(\delta_{ki}P_j -
%\delta_{kj}P_i\right) \otimes P_0
\;,\\
 &~~&  \cr
\lim_{c\to \infty}\;\Delta_{\theta^{0i}}(V_i)&=&\Delta_0(V_i) -
\frac{\hat{\theta}^{0k}}{\hat{\kappa}}\Pi_i \wedge
\Pi_k\;.\label{zcoppy100}
\end{eqnarray}
The algebraic relations (\ref{dlww2.1x}) and coproducts
(\ref{zcoppy1})-(\ref{zcoppy100}) constitute the canonical deformed
Galilei Hopf algebra $\,{\mathcal
U}_{\frac{\hat{\theta}}{\hat{\kappa}}}({\mathcal G})$ with the
corresponding  classical
$\texttt{r}_{\frac{\hat{\theta}}{\hat{\kappa}}}$-matrix
\begin{equation}
\texttt{r}_{\frac{\hat{\theta}}{\hat{\kappa}}}=\frac{\hat{\theta}^{0k}}{\hat{\kappa}}\,\Pi_{0}\wedge
\Pi_{k}\;, \label{hhgrmatix}
\end{equation}
and the following twist factor
\begin{equation}
{\mathcal{K}}_{\frac{\hat{\theta}}{\hat{\kappa}}}=\exp
\left(\frac{i\hat{\theta}^{0k}}{\hat{\kappa}}\Pi_{0 }\wedge
\Pi_{k}\right)\;. \label{fggfactor}
\end{equation}
Unfortunately, one can check that for $\theta^{ij} \ne 0$ the
contracted coproducts become in the limit $c\to \infty$ divergent.

If we   introduce a new deformation parameter $\overline{\kappa}$;
$\kappa = \overline{\kappa}c$
$(\left[\,\overline{\kappa}\,\right]=TL^{-2})$, one can repeat the
above contraction procedure. In such a way,  in the contraction
limit, we obtain however the undeformed Galilei Hopf algebra
$\,{\mathcal U}_{0}({\mathcal G})$.

In the case of  Lie-algebraic space-time noncommutativity
corresponding to $\,{\mathcal U}_{\zeta}({\mathcal P})$ with $\zeta
= \frac{1}{\kappa}$ the situation is more complicated. For
${\kappa}=\hat{\kappa}/c$ and carrier
 $\{\,M_{kl}, P_0\,\}$, we
get the Lie-algebraic Galilei algebra $\,{\mathcal
U}_{\hat{\kappa}}({\mathcal G})$ with classical algebraic sector
(\ref{nnn}). Its coproducts for momentum and rotation are given by
(\ref{ffdlww2.2})-(\ref{ffdlww2.22}) with $\gamma=0$, $\zeta =
\frac{1}{\hat{\kappa}}$, while for boost it looks as follows
\begin{eqnarray}
\Delta_\zeta(V_i)&=&\Delta_0(V_i)+\frac{1}{2{\hat{\kappa}}}K_{k
l}\wedge
 \Pi_i+i\left[V_i,K_{k l }\right]\wedge
\sin(\frac{1}{2{\hat{\kappa}}} \Pi_0 ) \notag \\
&+&\left[\left[%
V_i,K_{k l }\right],K_{k l }\right]\perp
(\cos(\frac{1}{2{\hat{\kappa}}}  \Pi_0  )-1)  \label{copp2} \\
&+&K_{k l }\sin(\frac{1}{2{\hat{\kappa}}} \Pi_0 )\perp
\frac{1}{2{\hat{\kappa}}} \left(\delta_{k i}\Pi_l - \delta_{l i} \Pi_k  \right) \notag \\
&-&\frac{1}{2{\hat{\kappa}}} \left(\delta_{k i}\Pi_k+\delta_{l i}
\Pi_l \right)\wedge K_{k l }(\cos(\frac{1}{2{\hat{\kappa}}} \Pi_0
)-1)\;.
  \notag
\end{eqnarray}
One can also check that its classical $r$-matrix has the form
\begin{equation}
r_{\hat{\kappa}}^{(k,l)}=\frac{1}{2\hat{\kappa}}\,\Pi_{0}\wedge K_{k
l}\;, \label{fgrmatix}
\end{equation}
and the twist element is given by
\begin{equation}
{\mathcal{K}}_{\hat{\kappa}}^{(k,l)}=\exp
\frac{i}{2\hat{\kappa}}(\Pi_{0}\wedge K_{k l})\;. \label{dgfactor}
\end{equation}
In the case $\{\,M_{kl}, P_\gamma\;;\;\gamma\ne k,l,0\,\}$ it is
easy to see that the coproducts for rotations do not have the
 contraction limits. The similar situation
appears for "boost-like" carrier $\{\,M_{k0}, P_l\;;\;k\ne l\,\}$ -
the rotation coproduct is divergent.

For $\kappa = \overline{\kappa}c$, by simple calculation one can
check that in the case of "space-like" carrier $\{\,M_{kl},
P_\gamma\;;\;\gamma\ne k,l\,\}$ we get  the undeformed Galilei
algebra $\,{\mathcal U}_{0}({\mathcal G})$, i.e. its contracted
$r$-matrix disappears ($r_{\overline{\kappa}}=0$) and the coproducts
remain primitive. For "boost-like" carrier $\{\,M_{k0}, P_l\;;\;k\ne
l\,\}$ situation is less-trivial - the contracted Galilei Hopf
algebra $\,{\mathcal U}_{\overline{\kappa}}({\mathcal G})$ is given
by the classical commutation relations (\ref{nnn}) and the following
coproducts
\begin{eqnarray}
 \Delta_{\overline{\kappa}}(\Pi_0)&=&\Delta _0(\Pi_0) +
\frac{1}{2{\overline{\kappa}}} \Pi_l \wedge \Pi_k\;,\label{coppy1}\\
 &~~&  \cr
\Delta_{\overline{\kappa}}(\Pi_i)&=&\Delta
_0(\Pi_i)\;\;\;,\;\;\;\Delta_{\overline{\kappa}}(V_i)=\Delta_0(V_i)\;,\label{cop0}\\
 &~~&  \cr
\Delta_{\overline{\kappa}}(K_{ij})&=&\Delta_0(K_{ij})+
\frac{i}{2{\overline{\kappa}}}\left[K_{ij},V_k\right]\wedge \Pi_l +
\frac{1}{2{\overline{\kappa}}}V_k \wedge(\delta_{il}\Pi_j
-\delta_{jl}\Pi_i)  \;, \label{coppy100}
\end{eqnarray}
By straightforward calculations one can check that the above quantum
group can be obtain by twist transformation of the classical Galilei
algebra with use of the following factor
\begin{equation}
{\mathcal{K}}_{\overline{\kappa}}^{(l,k)}=\exp
\frac{i}{2\overline{\kappa}}(\Pi_{l}\wedge V_k)\;. \label{dghfactor}
\end{equation}

%Let us summarize our results.

\section{Noncommutative Galilei space-times}

In  previous sections  we found by nonrelativistic contractions five
deformations of Galilei Hopf algebras. Two of them correspond to the
simplest contraction (\ref{contr2}) with $c$-independent parameter
of deformation (Section 3). They are given by:\\

{\bf i)} the deformed relations (\ref{dlww3v})-(\ref{zadruzny}) for
canonically deformed
$\,{\mathcal U}_{\theta}({\mathcal G})$,\\
%and by

{\bf ii)}  the deformed formulas
(\ref{ffdlww2.2})-(\ref{fifdlww2.22}) for Lie-algebraic type
deformation with carrier

~~~~$\{\,M_{kl}, P_\gamma\;;\gamma\ne k,l,0\,\}$ - $\,{\mathcal U}_{\zeta}({\mathcal G})$.\\
\\
We also get the quantum groups corresponding to the contraction
(\ref{contr2}) with the parameter $c$ hidden in $\kappa =
{\hat{\kappa}}/c$ (see Section 4). In the case of canonical
deformation we obtain:\\

{\bf iii)}  the quantum Galilei group $\,{\mathcal
U}_{\frac{\hat{\theta}}{\hat{\kappa}}}({\mathcal G})$ equipped with
the deformed coproducts (\ref{zcoppy1})-

~~~~-(\ref{zcoppy100}).\\
\\
For Lie-algebraic noncommutativity, in the case of carrier
$\{\,M_{kl}, P_0\,\}$, we get:\\

{\bf iv)} the deformed  coproducts
(\ref{ffdlww2.2})-(\ref{ffdlww2.22}) with
$\gamma =0$ supplemented by the formula (\ref{copp2}).\\
\\
Finally, if we perform the contraction (\ref{contr2}) with parameter
$\kappa = {\overline{\kappa}}c$ we have:\\

{\bf v)} the Galilei Hopf algebra $\,{\mathcal
U}_{{\overline{\kappa}}}({\mathcal G})$ with deformed coproducts
(\ref{coppy1})-(\ref{coppy100}) for $\{\,M_{k0}, P_l\;;\;$

~~~~$k\ne l\,\}$.\\

In this section we construct nonrelativistic space-times
corresponding to the Galilei quantum groups {\bf i)} - {\bf
v)}\footnote{The similar classification of twist factors and
corresponding superspaces in the case of superconformal algebra has
been proposed in \cite{japan}.}. They are defined as the quantum
representation spaces (Hopf modules) for quantum Galilei algebras,
with action of the deformed symmetry generators satisfying suitably
deformed Leibnitz rules \cite{bloch}, \cite{3a}, \cite{3b}. The
action of Galilei group on a Hopf module of functions depending on
space-time coordinates $(t,x_i)$ is given by
\begin{equation}
\Pi_{0}\rhd
f(t,\overline{x})=i{\partial_t}f(t,\overline{x})\;\;\;,\;\;\;
\Pi_{i}\rhd f(t,\overline{x})=i{\partial_i}f(t,\overline{x})\;,
\label{a1}
\end{equation}
and
\begin{equation}
K_{ij}\rhd f(t,\overline{x}) =i\left( x_{i }{\partial_j} -x_{j
}{\partial_i} \right) f(t,\overline{x})\;\;\;,\;\;\; V_i\rhd
f(t,\overline{x}) =it{\partial_i} \,f(t,\overline{x})\;,\label{dsf}
\end{equation}
while the $\star$-multiplication of arbitrary two functions  is
defined as follows
\begin{equation}
f(t,\overline{x})\star_{{\cdot}} g(t,\overline{x}):=
\omega\circ\left(
 \mathcal{K}_{\cdot}^{-1}\rhd  f(t,\overline{x})\otimes g(t,\overline{x})\right) \;.
\label{star}
\end{equation}
In the above formula $\mathcal{K}_{\cdot}$ denotes  twist factor
corresponding to a proper Galilei group and $\omega\circ\left(
a\otimes b\right) = a\cdot b$.

Hence, we have:\\

{\bf i)}
%of soft algebra $\,{\mathcal U}_{\theta}({\mathcal G})$
the twist factor (\ref{gfactor}) is represented by
\begin{equation}
{\mathcal{K}}_{\theta }=\exp \left(-\frac{i}{2}\theta
^{kl}\partial_{k }\wedge \partial_{l}\right)\;, \label{repgfactor}
\end{equation}

~~~~and using the formulas (\ref{star}), (\ref{repgfactor}) one can
check that  "time-like" commutators

~~~~vanish
\begin{equation}
[\,t,x_i\,]_{{\star}_{{\theta }}} = t{{\star}_{\theta }} x_i -
x_i{{\star}_{{\theta }}}t = 0
 \;, \label{stcom}
\end{equation}

~~~~while  "space-like" ones introduce a canonical type of
noncommutativity
\begin{equation}
[\,x_i,x_j\,]_{{\star}_{{\theta }}} = i\theta^{kl}\left (
\delta_{ik}\delta_{jl}-\delta_{jk}\delta_{il}\right) = 2i\theta^{ij}
 \;.
\label{stcom1}
\end{equation}

~~~~The relations (\ref{stcom}) and (\ref{stcom1}) describe the
nonrelativistic "magnetic" form of known

~~~~canonical deformation of Minkowski space coordinates.\\

{\bf ii)} For Lie-algebraic deformation $\,{\mathcal
U}_{\zeta}({\mathcal G})$ the twist factor (\ref{swarozyc}) is
represented on  Hopf

~~~~module as follows (see (\ref{a1}), (\ref{dsf}))
\begin{equation}
\mathcal{K}_{\zeta}=  {\rm \exp}
\,\left(-\frac{i}{2}\zeta\,\partial_{\gamma}\wedge ( x_{k
}{\partial_l} -x_{l }{\partial_k})\right)\;\;\;;\; \gamma \ne
k,l,t\;. \label{swarozyc1}
\end{equation}

~~~~Using (\ref{star}) and (\ref{swarozyc1}) we obtain the following
commutation relations
\begin{equation}
[\, x_{i },x_{j }\,] _{\star_{{\zeta }} }= i\zeta \delta_{\gamma j}(
 \delta _{k i }x_{l }- \delta_{l i }x_{k }) +i\zeta \delta_{\gamma
i}(\delta _{ l j }x_{k } - \delta _{k j }x_{l } )
\;\;\;,\;\;\;[\,t,x_i\,]_{{\star}_{{\zeta }}} =  0\;, \label{ssstar}
\end{equation}

~~~~which define the new model of Lie-algebraic $\zeta$-deformed
space-time for Galilei

~~~~algebra $\,{\mathcal U}_{\zeta}({\mathcal G})$.\\

%Let us now  turn to the algebras contracted with $c$-dependent
%deformation parameters.\\

{\bf iii)} In this case the twist factor (\ref{fggfactor}) is given
by
\begin{equation}
{\mathcal{K}}_{\frac{\hat{\theta}}{\hat{\kappa}}}=\exp
\left(-\frac{i\hat{\theta}^{0k}}{\hat{\kappa}}\partial_{t }\wedge
\partial_{k}\right)\;. \label{ssswarga}
\end{equation}

~~~~while corresponding space-time noncommutativity, due to
(\ref{star}) and (\ref{ssswarga}), has the

~~~~form
\begin{equation}
[\,t,x_i\,]_{{\star}_{{\frac{\hat{\theta}}{\hat{\kappa}}}}} =
2\frac{i\hat{\theta}^{0i}}{\hat{\kappa}}
\;\;\;,\;\;\;[\,x_i,x_j\,]_{{\star}_{{\frac{\hat{\theta}}{\hat{\kappa}}}}}
= 0\;, \label{swarga500}
\end{equation}

~~~~and corresponds to "electric" deformation of Minkowski space
coordinates.\\

{\bf iv)} For $\,{\mathcal U}_{{\hat \kappa}}({\mathcal G})$ the
twist factor (\ref{dgfactor}) and corresponding space-time are given
by
\begin{equation}
\mathcal{K}_{\hat {\kappa}}=  {\rm \exp} \,\left(-\frac{i}{2\hat
{\kappa}}\,\partial_{t}\wedge ( x_{k }{\partial_l} -x_{l
}{\partial_k})\right)\;, \label{swar2}
\end{equation}

~~~~and
\begin{equation}
[\, t,x_{i }\,]_{{\star}_{{\hat\kappa}}}= \frac{i}{{\hat\kappa}} (
\delta _{ li }x_{k }-\delta _{k i }x_{l } )\;\;\;,\;\;\;[\,x_{i
},x_j\,]_{{\star}_{{\hat\kappa}}} =  0 \;, \label{sistar}
\end{equation}

~~~~respectively. We see that we obtained by (\ref{sistar}) the
nonrelativistic variant of

~~~~$\kappa$-Minkowski space with classical three-space and quantum time. \\

{\bf v)} Finally, for quantum group $\,{\mathcal
U}_{\overline{\kappa}}({\mathcal G})$ we have (see
(\ref{dghfactor}))
\begin{equation}
\mathcal{K}_{\overline{\kappa}}=  {\rm \exp}
\,\left(-\frac{i}{2\overline{\kappa}}\,\partial_{l}\wedge
t{\partial_k} \right)\;. \label{swar1}
\end{equation}

~~~~We also obtain the following commutation relations
\begin{equation}
[\, x_{i },x_{j }\,] _{\star_{\overline{\kappa}}}=
\frac{i}{\overline{\kappa}}t(\delta _{l i }\delta_{k j }-\delta _{
ki }\delta _{l j
})\;\;\;,\;\;\;[\,t,x_i\,]_{\star_{\overline{\kappa}}} = 0 \;.
\label{ysesstar}
\end{equation}

~~~~The deformation (\ref{ysesstar}) of nonrelativistic space-time
describes conceptually new type

~~~~of noncommutativity. It should be interesting to consider
dynamical models based

~~~\,on deformed space-time (\ref{ysesstar}), with quantum space and
classical time coordinate

~~~\,(see \cite{cm}).

\section{Final remarks}

In this paper we investigate three different nonrelativistic
contractions of soft and Lie-algebraic Poincar\'{e} algebras. In
such a way we get five  deformations of Galilei groups - three of
Lie-type and two of a soft kind. Further, we  derive as well the
corresponding nonrelativistic space-times representing the Hopf
module of our Galilean quantum algebras.

The present studies can be extended in various ways.  First of all,
one can construct associated with the presented algebras suitable
phase spaces derived with the help of Heisenberg double construction
\cite{majid}, \cite{majid1}, \cite{majid3}, \cite{piotr}. Besides,
we could also look for the dual Galilei quantum groups, which can be
recovered with use of FRT method (see \cite{frt}) or by canonical
quantization of a proper Poisson-Lie structure \cite{kowclas},
\cite{Poistru}. Finally, one can  ask about explicit realization
 of others
$D=3+1$  Galilei Hopf algebras and their  deformed space-times,
which are  predicted by the general classification of all
"nonrelativistic" classical r-matrices \cite{kowclas} (see also for
two-dimensional case
\cite{lodz1}, \cite{lodz3}). %These
% problems are under consideration.
These problems are  under considerations and they are postponed for
further investigation.

\section*{Acknowledgments}
The author would like to thank  Jerzy Lukierski and  Mariusz
Woronowicz for many valuable discussions.

\end{document}